\documentclass[twocolumn,aps,prd,10pt,superscriptaddress]{revtex4-1}

\usepackage{amsmath}
\usepackage{amssymb}
\usepackage{amsfonts}
\usepackage{graphicx}

\newcommand{\etal}{{\it et~al}.}
\newcommand{\rhoc}{\rho_\text{c}}
\newcommand{\ER}{E_\text{R}}
\newcommand{\nn}{\boldsymbol{n}}
\newcommand{\av}[1]{\langle#1\rangle}
\newcommand{\LL}{{\mathcal L}}

\newcommand{\DD}{\mathcal{D}}

\newcommand{\dd}{\text{d}}
\newcommand{\TT}{\text{TT}}
\newcommand{\ppp}{\\[3pt]}
\newcommand{\nppp}{\nonumber\\[3pt]}

\newcommand{\bra}[1]{\langle#1|}

\newcommand{\ket}[1]{|#1\rangle}
\newcommand{\Bra}[1]{(#1|}

\newcommand{\Ket}[1]{|#1)}
\newcommand{\hc}{\text{H.c.}}
\newcommand{\kk}{\boldsymbol{k}}
\newcommand{\kb}{\mathbf{k}}
\newcommand{\rr}{\boldsymbol{r}}
\newcommand{\ngw}{N_{\text{gw}}}
\newcommand{\nvac}{N_{\text{vac}}}
\newcommand{\vac}{\text{vac}}

\newcommand{\gw}{\text{gw}}
\newcommand{\intt}{\text{int}}
\newcommand{\tr}{\text{Tr}}

\newcommand{\Sec}[1]{Sec.~\ref{#1}}
\newcommand{\apx}[1]{appendix~\ref{#1}}
\newcommand{\Fig}[1]{Fig.~\ref{#1}}
\newcommand{\Figg}[2]{Figs.~\ref{#1} and \ref{#2}}
\newcommand{\Ref}[1]{Ref.~\cite{#1}}

\newcommand{\refs}[1]{~\cite{#1}}
\newcommand{\eq}[1]{~\eqref{#1}}
\newcommand{\eqq}[2]{~\eqref{#1} and \eqref{#2}}
\newcommand{\Eq}[1]{Eq.~\eqref{#1}}
\newcommand{\Eqq}[2]{Eqs.~\eqref{#1} and \eqref{#2}}

\newcommand{\Eqqqq}[4]{Eqs.~\eqref{#1}, \eqref{#2}, \eqref{#3}, and \eqref{#4}}

\newcommand{\Eqs}[1]{Eqs.~\eqref{#1}}
\newcommand{\Idash}{I\hspace{-4.6pt}\raisebox{1.5pt}{-\hspace{-2pt}-}}

\begin{document}


\title{Quantum principle of sensing gravitational waves: From the zero-point\\ fluctuations to the cosmological stochastic background of spacetime}

\author{Diego A. Qui\~nones}
\email{pydaqo@leeds.ac.uk}
\affiliation{School of Physics and Astronomy, University of Leeds, Leeds LS2 9JT, United Kingdom}

\author{Teodora Oniga}
\email{t.oniga@abdn.ac.uk}
\affiliation{Department of Physics, University of Aberdeen, Aberdeen AB24 3UE, United Kingdom}

\author{Benjamin T. H. Varcoe}
\email{b.varcoe@leeds.ac.uk}
\affiliation{School of Physics and Astronomy, University of Leeds, Leeds LS2 9JT, United Kingdom}

\author{Charles H.-T. Wang}
\email{Corresponding author, c.wang@abdn.ac.uk}
\affiliation{Department of Physics, University of Aberdeen, Aberdeen AB24 3UE, United Kingdom}


\begin{abstract}
\vspace{10pt}
We carry out a theoretical investigation on the collective dynamics of an ensemble of correlated atoms, subject to both vacuum fluctuations of spacetime and stochastic gravitational waves. A general approach is taken with the derivation of a quantum master equation capable of describing arbitrary confined nonrelativistic matter systems in an open quantum gravitational environment. It enables us to relate the spectral function for gravitational waves and the distribution function for quantum gravitational fluctuations and to indeed introduce a new spectral function for the zero-point fluctuations of spacetime. The formulation is applied to two-level identical bosonic atoms in an off-resonant high-$Q$ cavity that effectively inhibits undesirable electromagnetic delays, leading to a gravitational transition mechanism through certain quadrupole moment operators. The overall relaxation rate before reaching equilibrium is found to generally scale collectively with the number $N$ of atoms. However, we are also able to identify certain states of which the decay and excitation rates with stochastic gravitational waves and vacuum spacetime fluctuations amplify more significantly with a factor of $N^2$. Using such favourable states as a means of measuring both conventional stochastic gravitational waves and novel zero-point spacetime fluctuations, we determine the theoretical lower bounds for the respective spectral functions. Finally, we discuss the implications of our findings on future observations of gravitational waves of a wider spectral window than currently accessible. Especially, the possible sensing of the zero-point fluctuations of spacetime could provide an opportunity to generate initial evidence and further guidance of quantum gravity.
\vspace{30pt}
\end{abstract}


\maketitle

\section{Introduction}
\label{sec:intr}

\subsection{Background}

Recently, there has been considerable interest in experimenting with gravity using quantum instrument\refs{Amelino2013, Pfister2016}, to better explore its theory, predictions, and applications in uncontested regimes\refs{Wang2006a, Kiefer2013, Oniga2016a}. This trend has been established through laboratory measurements of the gravitational acceleration by ultracold atom interferometers with accuracies exceeding that achievable by conventional gravimeters\refs{Muller2008}. The realization of such quantum gravity and gravity gradient sensors as portable devices would herald a new level of quantum applications in engineering and technology\refs{Carraz2014}.

For physicists, it has been hoped that major progress across a range of areas could be enabled by the availability of precision quantum gravity sensors. These include testing of the equivalence principle as a foundation of general relativity in the quantum domain\refs{gauge2009}, confronting macroscopic quantum behaviour with a microgravity environment\refs{Muntinga2013}, and assessing a long range effect such as gravitational time dilation on quantum coherence\refs{Pikovski2015}. Above all, quantum experiments may promise fundamental breakthroughs by offering a unique opportunity to access phenomena from quantum gravity and Planck-scale physics that would otherwise have required an inconceivable $10^{19}$ GeV Grand Unified Theory (GUT)-scale particle collider\refs{Amelino2000, Wang2006, Lamine2006, Everitt2013}.

There are a number of counterintuitive effects unavailable in the classical domain to justify novelties and advantages of quantum measurements. For Bell's inequality related experiments\refs{PPT1, Handsteiner2017}, quantum entanglement and nonlocality are the key. More relevant to precision measurements, in the case of atom interferometry metrology, the short matter wavelength underpins a fine scale resolution. When atom interferometry is applied to gravity measurements including gradiometry and spacetime fluctuations, the slowness of the motion for cold atoms contributes to a longer interaction time to build up effects in terms of phase shift due to acceleration and decoherence due to granulation\refs{gauge2009}.

For quantum systems with a large number of correlated particles, nontrivial collective behaviours can arise\refs{Oniga2016b, Oniga2016c}. An early remarkable example shows up in Hanbury Brown and Twiss's intensity interferometry\refs{HBT1956}. Collective behaviours can give rise to significant amplifications of quantum effects associated with a large number of correlated particles, as exemplified in Dicke's seminal superradiance\refs{Dicke1954}. In this paper, we will analyze previously unexploited collective quantum interactions between a large number of particles with gravitational waves that may be relevant for their detections beyond the existing observation windows e.g. in the frequency domain of fundamental, astronomical and cosmological interest\refs{LIGO2009, LIGO2016b, GWC2016, SGWB2016}.

In particular, we consider stochastic gravitational waves of frequencies higher than the existing detection range, that may have a primordial origin and carry imprints of postinflation structure formation processes\refs{LIGO2009, kiefer2000, Easther2006, Oniga2017}. The nature and possible observation of such stochastic gravitational waves have attracted significant recent attention since the groundbreaking detection of gravitational waves by the LIGO and Virgo collaborations\refs{LIGO2016} with the anticipation of broader gravitational wave astronomy and cosmology\refs{GWC2016, Schutz1999, Sesana2016}.
%
%
In what follows, we present a theoretical analysis with Gaussian stochastic gravitational waves\refs{Allen1996} having a generic distribution function $\ngw(\omega)$ and the equivalent spectral function $\Omega_\gw(f)$. The atoms are assumed to be bosonic so that the collective quantum  dynamics of a large number of atoms occupying a common state is possible.

The effects of spacetime curvature on individual atoms have been considered in a number of works\refs{Fischbach1981, Fischer1994, Marques2002, Olmo2008, Zhao2007}, showing slight shifts in the energy spectrum and weak transitions between the energy levels of the atoms induced by gravitational interactions. Gravity has been proposed to induce decoherence in atomic states\refs{Blencowe2013, Oniga2016a, Quinones2016}, making the analysis of the evolution of atomic states a potential method for the detection of gravitational fluctuations\refs{Parker1980, Parker1982, Leen1983, Pinto1995}. However, these effects have also been found to be generally too small to be measured in practice\refs{Fischer1994}.

It can be noted that Rydberg atoms\refs{Gallagher2008, Saffman2010} have received particular attention for atom-gravity interactions due to a number of their distinct physical characteristics they possess: Their large principal quantum number $n$ means that the effective mass quadrupole moment, which provides external gravitational wave coupling and scales with $n^4$, can be greatly enhanced. These atoms are very stable having extremely long lifetimes of the internal transitions up to the order of 1 second allowing us to neglect them in many practical situations\refs{Branden2010}. Furthermore, increasingly large numbers of Rydberg atoms with high principal quantum numbers $n>10^2$ have been produced in laboratories\refs{Ye2013}. The investigations of Rydberg atoms as a many-body system have often featured in the recent literature\refs{Browaeys2013} with their collective quantum dynamics receiving current research interest both theoretically\refs{Jaksch2000, Lukin2001, Olmos2010, Lee2012} and experimentally\refs{Gaetan2009, Wilk2010, Heidemann2007, Dudin2012, Mourachko1998, Schauss2012, Kumar2016}.

\subsection{Summary}

Rather than considering gravitational interactions with individual atoms, here we investigate the coherent quantum interactions for a large number of correlated atoms with gravitational waves. We find that the resulting collective interactions can lead to amplified quadrupole transition rates of atomic states in a fashion akin to superradiance\refs{Dicke1954}.

\begin{figure}[!ht]
\includegraphics[width=1\linewidth]{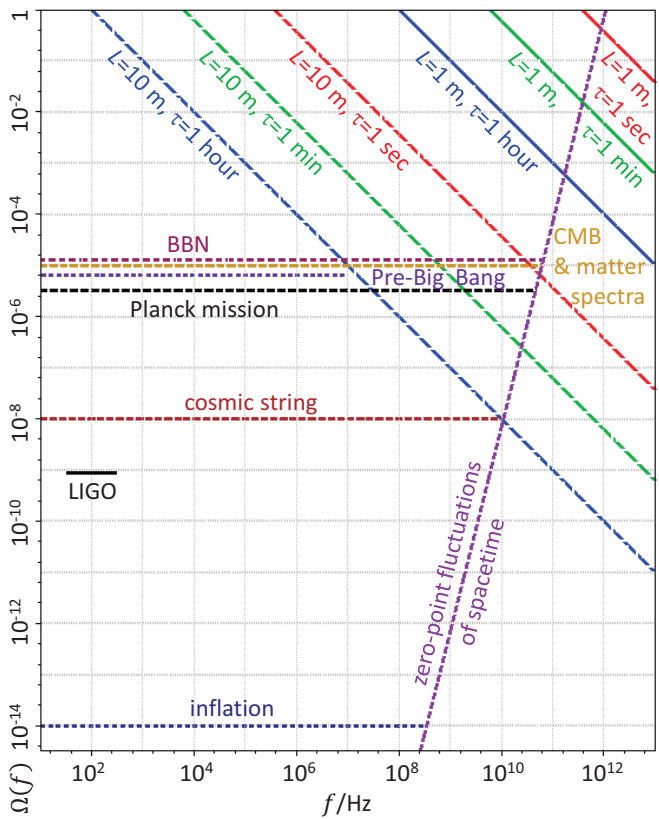}
\caption{
Theoretical lower bounds for the spectral functions of stochastic gravitational waves $\Omega(f)$ determined by \Eq{Omf2} that could be detected using an ensemble of correlated two-level atoms contained in an off-resonant high-$Q$ cavity of volume $L^3$ and measurement time $\tau$. Specifically, we consider heliumlike atoms with atomic quantum numbers $n_1=n_2=2$, $l_1=l_2=1$ and $-m_1=m_2=1$, with a transition frequency induced by a Zeeman-type shift.
For conventional stochastic gravitational waves, we have $\Omega(f)=\Omega_\gw(f)$ with $\tau=\tau_\gw$, whereas for zero-point, i.e. vacuum, metric fluctuations, we have $\Omega(f)=\Omega_\vac(f)$ with $\tau=\tau_\vac$ introduced in this work by \Eqq{tt}{Omvac}. In both cases, we choose $\tau=1$ sec, min, and h as possible (future) transition times for the atoms with $L=$ 1 and 10 m in generating the above diagram.
It suggests, for example, measuring the zero-point fluctuations of spacetime would require a 1 m$^3$ cavity with a measurement time of 1 sec at $f=1$ THz approximately. Note that regions where the values of $\Omega_\gw(f)$ are below that of $\Omega_\vac(f)$ would present a less stochastic gravitational wave signal to vacuum fluctuations ``noise'' ratio and hence the detection of the latter would be more likely using the discussed quantum methods. For comparison, we have superimposed the expected spectral functions of stochastic gravitational waves from various sources such as inflation, up to frequencies dominated by spacetime fluctuations, that may be detected by different methods such as LIGO from \Ref{LIGO2009} and its references.
}
\label{fig1}
\end{figure}

This collective amplification could be compared and combined with the application of Rydberg atoms, which are more sensitive as individual atoms to spacetime perturbations\refs{Pinto1995, Pinto1993}. However, their larger size can also decrease the number density and increase dipole-dipole interactions of the atoms. Consequently, we find that smaller atoms can provide more collective strength for gravitational wave interactions as described below, with specific numerical examples shown in \Fig{fig1}, using an ensemble of heliumlike atoms for detecting high frequency stochastic gravitational waves and spacetime fluctuations.
Our investigations may form a basis for a possible future detection scheme for gravitational waves with a wider spectral window, especially towards higher frequencies than currently accessible, which may be of fundamental and cosmological importance.

In the present theoretical investigation, we hypothesize for simplicity a setup of $N$ identical two-level atoms with a quadrupole moment transition frequency $f=\omega/2\pi$ used to evaluate the distribution function $\ngw(\omega)$ and the equivalent spectral function $\Omega_\gw(f)$ for the stochastic gravitational waves. Further refinement is possible to reduce the present level of idealization but will be deferred to future work. The system of atoms is considered to be placed in an off-resonant high-$Q$ cavity with controlled boundary conditions that effectively inhibits undesirable electromagnetic delays\refs{Kleppner1981, Kleppner1985, Varcoe2006}.
The presence of the high-$Q$ cavity can also change the quantum electrodynamics of the atom to modify the interaction between the nucleus and the electron so as to increase the quadrupole interaction while suppressing the dipole-dipole interactions between the atoms \refs{Marrocco1998}.

\begin{figure}[!ht]
\vspace{12pt}
\includegraphics[width=1\linewidth]{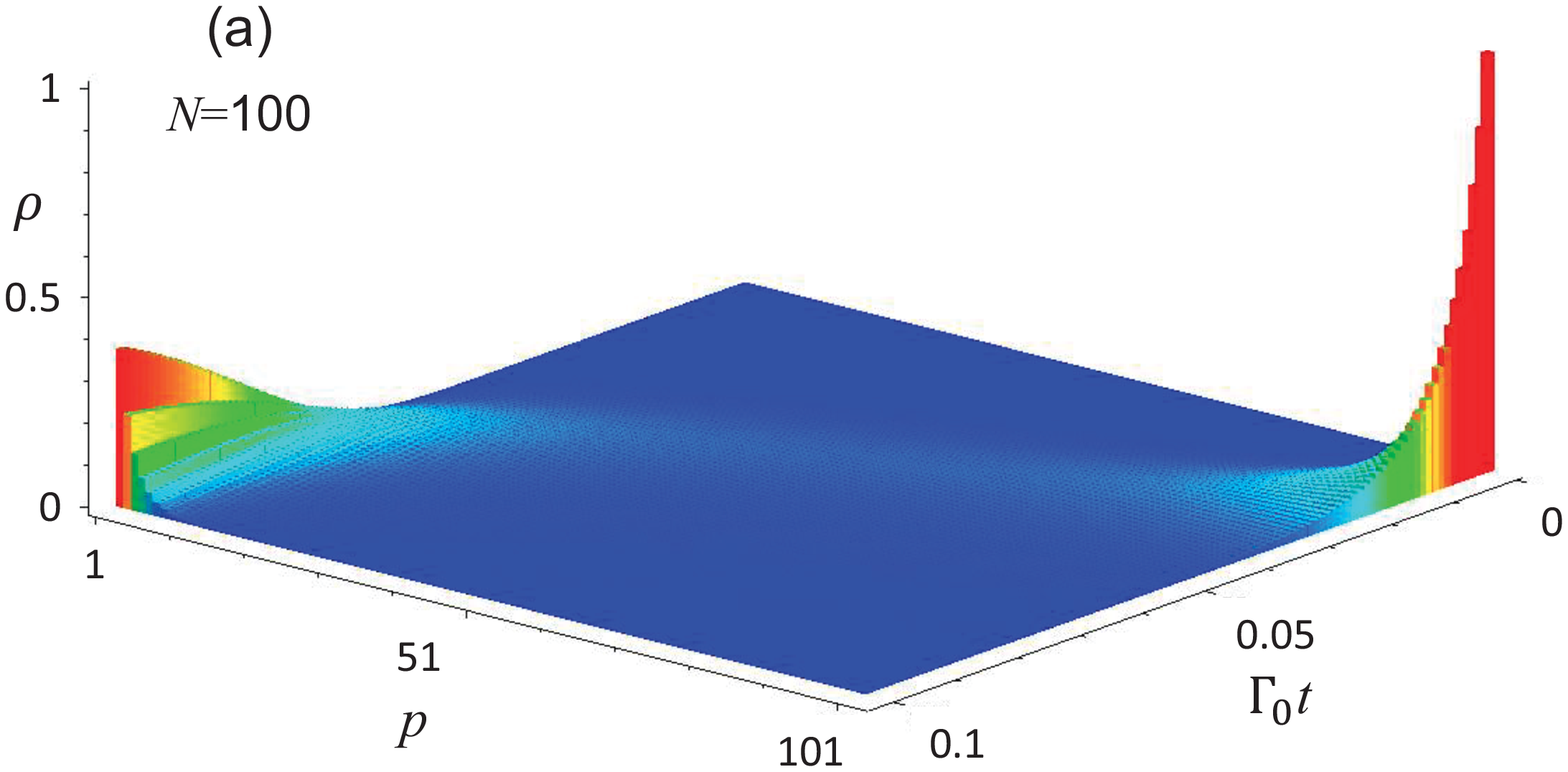}\\
\vspace{20pt}
\includegraphics[width=1\linewidth]{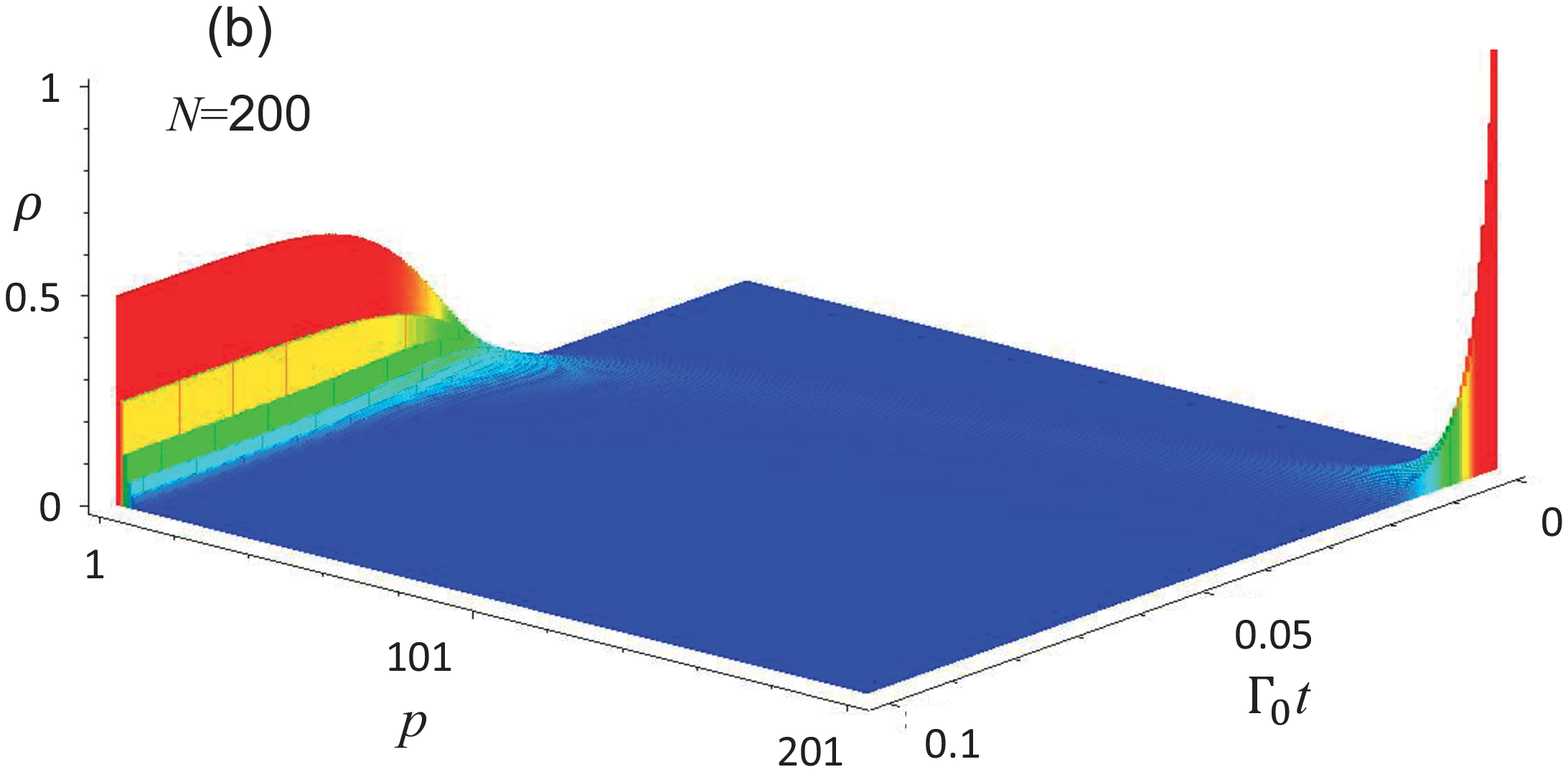}
\vspace{5pt}
\caption{Simulations with $N$ atoms using the master equation\eq{maseqn} with the quantum dissipator\eq{diss5} for the correlated two-level atoms and negligible matter interaction $H_\intt$.
From \Eq{diss7} the relaxation time is approximately given by
$\tau_N=(N\Gamma_0)^{-1}$.
The initial state of the forms\eqq{NN}{pp} at $t=0$ is given by $\ket{0,N}=\ket{p=N+1}$ with all $N$ atoms occupying the excited atomic level. During the dynamical evolution, the density matrix remains diagonal of the form\eq{dia} and relaxes into an inverse exponential profile represented by \Eq{rs}, yielding the asymptotic height of
$\rho_{p=1}=[\ngw(\omega_0)(1-e^{-N/\ngw(\omega_0)})]^{-1}$.
The stochastic gravitational wave distribution function evaluated at the atomic transition frequency $\omega_0$ is chosen to be $\ngw(\omega_0)=1$.
Accordingly, in plot (a) with $N=100$ atoms, the initial state corresponding to the spike of height 1 at $(p,\Gamma_0 t)=(101,0)$ relaxes into an inverse exponential profile peaked at
$(p,\Gamma_0 t)=(1,0.1)$ on time scale of $\tau_N=0.01/\Gamma_0$.
Similarly, in plot (b) with $N=200$ atoms, the initial state corresponding to the spike of height 1 at $(p,\Gamma_0 t)=(201,0)$ relaxes into an inverse exponential profile similarly peaked at
$(p,\Gamma_0 t)=(1,0.1)$ but on a halved time scale of $\tau_N=0.005/\Gamma_0$.
}
\label{fig2}
\end{figure}

In \Sec{sec:mas}, we further the theoretical foundation of how general confined nonrelativistic matter interacts with an open gravitational environment, relevant to, for example, a system of atoms in a cavity attracting much recent attention. Specifically, we provide a general master equation through \Eq{diss3} with negligible self-interactions to focus on the collective interactions between such a matter system and gravitational waves in the environment due to both vacuum and stochastic fluctuations.

In \Sec{sec:trans}, the theoretical formulation is applied to a two-level atom that allows us to derive specific transition mechanisms through the quadrupole moment operator with respect to the two atomic states given by \Eq{diss4}, satisfying a set of  selection rules\eq{selec}. In particular, we derive a formula for the gravitational transition rate $\Gamma_0$ in \Eq{gg0} for such a two-level atom. Unsurprisingly, the transition rate $\Gamma_0$ for a single atom is too small to be measured under typical laboratory conditions, as has been pointed out even for Rydberg atoms\refs{Fischer1994}.

In \Sec{sec:2level}, we invoke the generality of the formalism developed in \Sec{sec:mas} to extend the interaction between a two-level atom and gravity to accommodate a correlated ensemble of $N$ such atoms. In this case, as illustrated in \Fig{fig2}, we find an overall factor of $N$ collective amplification of the transition rate from any initial state to the final equilibrium state satisfying the distribution in the large-$N$ limit\eq{rs}.

\begin{figure}[!ht]
\includegraphics[width=1\linewidth]{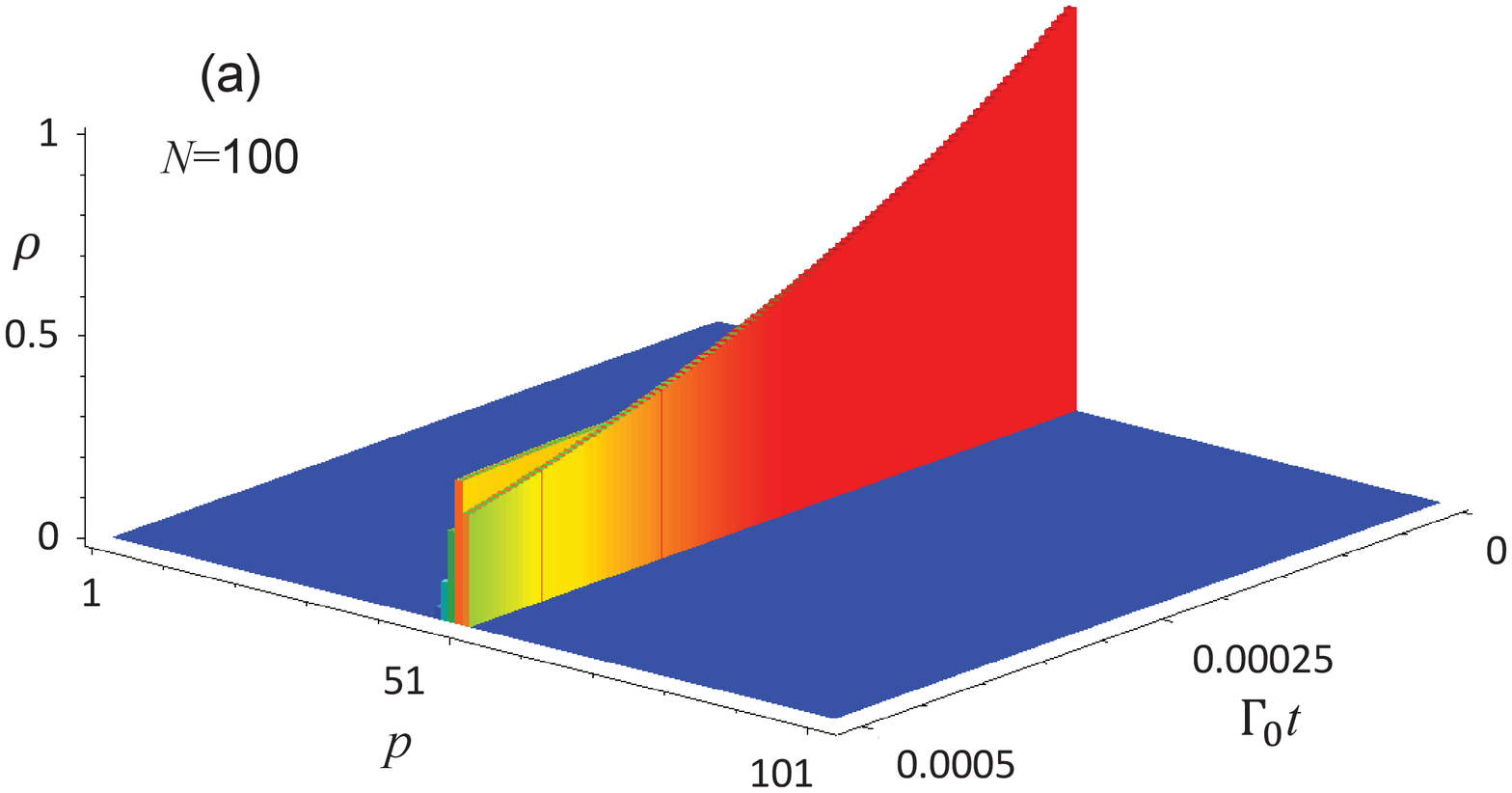}\\
\vspace{20pt}
\includegraphics[width=1\linewidth]{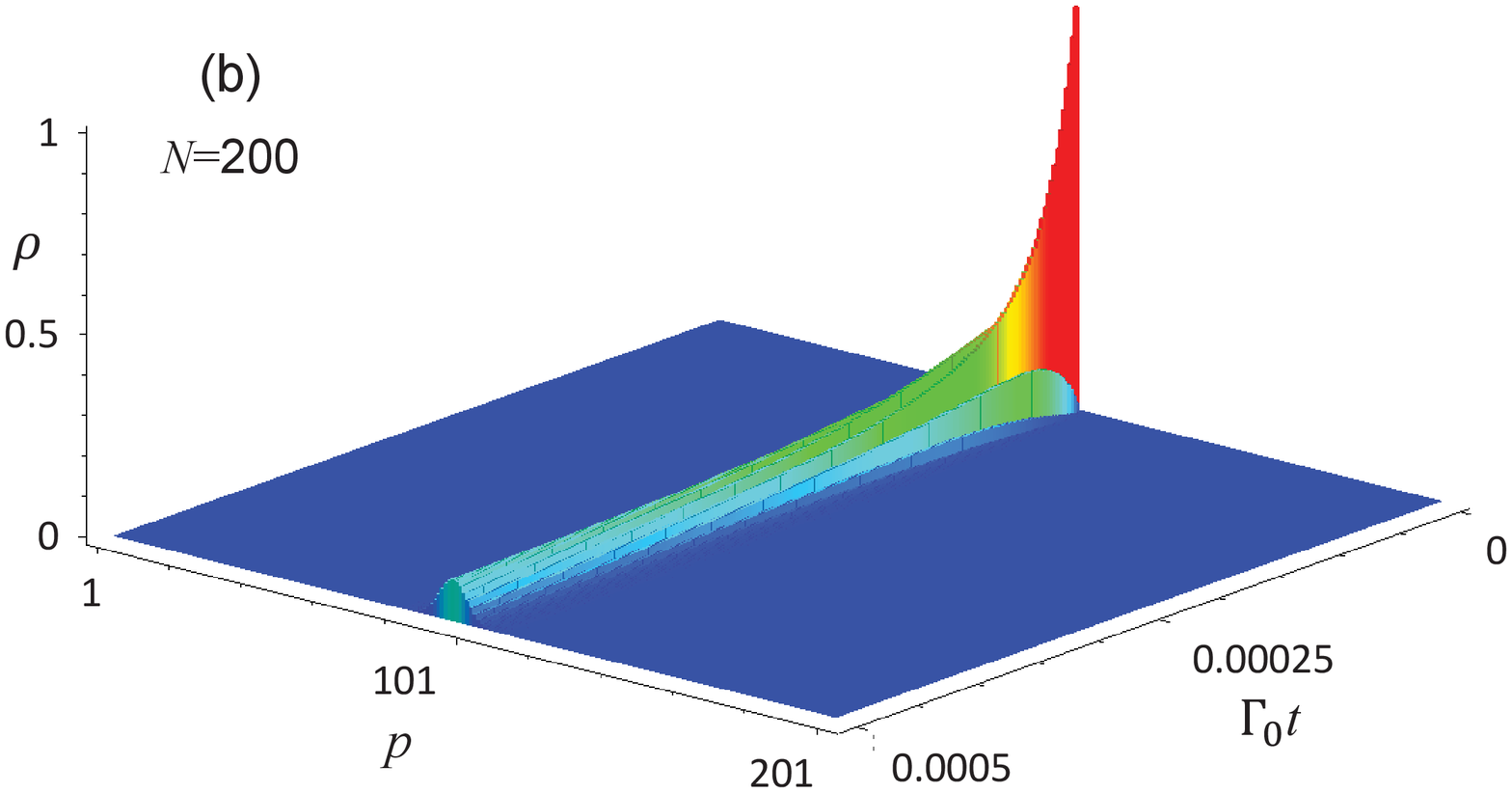}
\caption{Simulations of short-time collective spontaneous decays due to vacuum fluctuations of spacetime without additional stochastic gravitational waves so that $\ngw(\omega_0)=0$.
In plot (a) with $N=100$ atoms, the initial state corresponds to the spike of height 1 at $(p, t)=(50,0)$. As shown, it decays into adjacent lower states with a relaxation time $\tau_\vac=4/({N^2\Gamma_0})=0.0004/\Gamma_0$ according to \Eqq{ggvac}{tt}.
In plot (b) with $N=200$ atoms, the initial state corresponds to the spike of height 1 at $(p, t)=(100,0)$. For a doubled $N$, this state decays into adjacent lower states with a quarter of the preceding relaxation time with
$\tau_\vac=4/({N^2\Gamma_0})=0.0001/\Gamma_0$.}
\label{fig3}
\end{figure}

In \Sec{sec:gw}, we establish a relation between the spectral function $\Omega_\gw(f)$ for gravitational waves and distribution function $\ngw(\omega)$ for open quantum gravitational systems and introduce a new spectral function $\Omega_\vac(f)$ for the zero-point, i.e. quantum vacuum, fluctuations of spacetime.
Although we have introduced a spectral function for vacuum fluctuations of spacetime $\Omega_\vac(f)$ in \Eq{Omvac}, it should be emphasized that the detection of zero-point fluctuations of spacetime requires a quantum detector through decoherence as classical detectors have already decohered and so can no longer provide this function.

Indeed, the possibility of sensing the zero-point fluctuations of spacetime provides a unique physical characteristics for the collective quantum process which could lead to initial evidence and further guidance of quantum gravity.
For optimal measurements, we have identified favourable states of which the decay and excitation rates with gravitational waves and spacetime fluctuations scale amplify with a factor of $N^2$, as described by \Eqq{gggw}{ggvac} and illustrated in \Figg{fig2}{fig3}.

We then determine the theoretical lower bounds given by \Eq{Omf2} for the spectral functions of gravitational waves that could in principle be detected using an ensemble of identical two-level atoms given their excitation states and the size of the cavity as well as measurement time. The lower bound relation\eq{Omf2} applies to both conventional stochastic gravitational waves and novel zero-point spacetime fluctuations. Importantly, we find that, although as individual atoms with larger quadrupole moments, Rydberg atoms may have stronger interactions with gravitational waves, smaller atoms can nonetheless provide more collective strength for such interactions thanks to a higher possible number density of the atoms in a cavity.
Specific numerical examples using an ensemble of heliumlike atoms are studied, leading to new prospects for detecting high frequency stochastic gravitational waves and spacetime fluctuations as shown in \Fig{fig1}.

Finally, in \Sec{sec:conl}, we conclude this work with an additional discussion.

\section{Gravitational master equation for confined nonrelativistic systems}
\label{sec:mas}

In the interaction picture with relativistic units where $c=1$, the general gravitational master equation for  matter that may be interacting and relativistic has recently been obtained in \Ref{Oniga2016a} to be
\begin{eqnarray}
\dot\rho
&=&
-\frac{i}{\hbar} [H_\intt, \rho]
+\DD\rho
\label{maseqn}
\end{eqnarray}
with $H_\intt$ describing matter interaction with itself or other fields not part of the gravitational environment,  and the non-Markovian quantum dissipator
\begin{eqnarray}
\DD\rho
&=&
-\frac{8\pi G}{\hbar}
\int\! \frac{\dd^3 k}{2(2\pi)^3k}
\,\times
\nppp&&
\hspace{-10pt}
\Big \{
\int_{0}^{t} \!\dd t'
e^{-i k (t - t')}
\big(
[
\tau^\dag_{ij} (\kk, t),\,
\tau_{ij}(\kk, t') \rho
]
\nppp&&
\hspace{-10pt}
\,+\,\ngw(\omega_k)\,
[
\tau^\dag_{ij} (\kk, t),\,
[
\tau_{ij}(\kk, t'),\,
\rho
]]
\big)
+\hc
\Big\}
\label{diss}
\end{eqnarray}
describing in general nonunitary statistical quantum evolution of the matter system as a result of dissipative iterations with the environment \refs{Oniga2017, Oniga2016a}. Here, $\ngw(\omega)$ is a general distribution function for the fluctuating gravitational  environment assumed to be Gaussian. If this environment is in thermal equilibrium at temperature $T$ then $\ngw(\omega)$ is given by the Planck distribution\begin{eqnarray}
\ngw(\omega)=\frac1{e^{\hbar \omega/k_B T}-1}.
\label{Pdistgw}
\end{eqnarray}
However, we will be considering generic stochastic gravitational waves described by $\ngw(\omega)$ not restricted to the Planck distribution\eqref{Pdistgw}.

Since typical compact nonrelativistic systems have spatial extensions much smaller than the effectively coupled gravitational wavelength\refs{Flanagan2005, MTW1973}, we have $e^{-i \kk\cdot\rr}\approx 1$ and therefore
\begin{eqnarray}
\tau_{ij}(\kk, t)
&=&
P_{ijkl}(\kk)
\int\!
\tau_{kl}(\rr, t)\,
e^{-i \kk\cdot\rr}\,\dd^3x
\nppp
&\approx&
\frac12
\ddot\Idash_{ij}^\TT(\kk,t)
\label{ITT1}
\end{eqnarray}
where
\begin{eqnarray}
\Idash_{ij}^\TT(\kk,t)
&=&
P_{ijkl}(\kk)\Idash_{kl}(t)
\label{ITT2}
\end{eqnarray}
and
\begin{eqnarray}
\Idash_{ij}(t)
&=&
\frac{1}{3}\,\int\dd^3x\,Q_{ij}(\rr)\,T^{00}(\rr,t)
\label{qij2}
\end{eqnarray}
in terms of the transverse-traceless (TT) projection operator $P_{ijkl}$\refs{Flanagan2005, Oniga2016a} and the quadrupole moment tensor
\begin{eqnarray}
Q_{ij}
&=&
3x^i x^j - \delta_{ij}\, r^2 .
\label{Qij}
\end{eqnarray}

It is useful to introduce the frequency domain reduced quadrupole moments $\Idash_{ij}(\omega)$ for some positive frequencies $\omega$ by writing
\begin{eqnarray}
\Idash_{ij}(t)
&=&
\sum_{\omega}
\big[
\Idash_{ij}(\omega)\, e^{-i\omega t} + \Idash_{ij}^\dag(\omega)\, e^{i\omega t}
\big].
\label{ITT3}
\end{eqnarray}
Let us consider the following integral in \Eq{diss},
\begin{eqnarray}
\lefteqn{
\int_{0}^{t} \!\dd t'
e^{-i k (t - t')}
\tau_{ij}(\kk, t')
}
\nppp
&=&
-\sum_{\omega}\frac{\omega^2}{2}\,\Idash_{ij}^\TT(\omega,\kk)
\int_{0}^{t} \!\dd t'
e^{-i k (t - t')}
e^{-i\omega t'}
\nppp&&
-\sum_{\omega}\frac{\omega^2}{2}\,\Idash_{ij}^{\TT\dag}(\omega,\kk)
\int_{0}^{t} \!\dd t'
e^{-i k (t - t')}
e^{i\omega t'}
\label{iet}
\end{eqnarray}
which is nonlocal in time representing the non-Markov memory effect.

To apply the Markov approximation, valid for physical time scales where short-term transient memories are damped away\refs{Breuer2002}, we change the time variable from $t'$ to $s$ using $t'=t-s$ in the above to get
\begin{eqnarray*}
\int_{0}^{t} \!\dd t'
e^{-i k (t - t')}
e^{\pm i\omega t'}
&\approx&
e^{\pm i\omega t}
\int_{0}^{\infty} \!\dd s\,
e^{-i (k\pm\omega) s}
\nppp
&=&
e^{\pm i\omega t}\Big[\,
\pi \delta(k\pm\omega) - i \,\text{P} \frac{1}{k\pm\omega}
\,\Big]
\end{eqnarray*}
where the Sokhotski-Plemelj theorem [see \Eq{fma} below] has been used in the last step.

Using the above and neglecting the the Cauchy principal values, as they do not contribute to quantum decoherence and dissipation, we see that \Eq{iet} becomes
\begin{eqnarray}
\lefteqn{
\int_{0}^{t} \!\dd t'
e^{-i k (t - t')}
\tau_{ij}(\kk, t')
}
\nppp
&=&
-\sum_{\omega}
\frac{\pi\omega^2}{2}\,\Idash_{ij}^\TT(\omega,\kk)\,
e^{-i\omega t} \delta(k-\omega)
\nppp&&
-\sum_{\omega}
\frac{\pi\omega^2}{2}\,\Idash_{ij}^{\TT\dag}(\omega,\kk)\,
e^{i\omega t} \delta(k+\omega)
\label{iet2}
\end{eqnarray}
where the second term does not contribute as $k \ge 0$ and $\omega > 0$.
Using \Eqs{ITT1} and \eqref{ITT3}, we have
\begin{eqnarray}
\lefteqn{
\tau_{ij}^\dag(\kk, t)
\approx
\frac12
\ddot\Idash_{ij}^{\TT\dag}(\kk,t)
}
\nppp
&=&
-\frac{\omega^2}2\big[
\Idash_{ij}^\TT(\omega,\kk)\, e^{-i\omega t}
+
\Idash_{ij}^{\TT\dag}(\omega,\kk)\, e^{i\omega t}
\big] .
\label{ITT4}
\end{eqnarray}
Substituting \Eqq{iet2}{ITT4} into the dissipator~\eqref{diss} and applying the rotating wave approximation we have
\begin{eqnarray}
\DD\rho
&=&
-\frac{G\omega^5}{8\pi \hbar}
\int\!{\dd\Omega(\kk)}
\Big \{
\big(
[
\Idash_{ij}^{\TT\dag}(\omega,\kk),\,
\Idash_{ij}^\TT(\omega,\kk)
\rho
]
\nppp&&
\hspace{-10pt}
+\ngw(\omega)\,
[
\Idash_{ij}^{\TT\dag}(\omega,\kk),\,
[
\Idash_{ij}^\TT(\omega,\kk),\,
\rho
]]
\big)
+\hc
\Big\}
\nppp
&=&
\frac{G\omega^5}{4\pi \hbar}
\int\!{\dd\Omega(\kk)}
P_{ijkl}(\kk)\times
\nppp&&
\hspace{-10pt}
\Big\{
(\ngw(\omega)+1)
\big[
\Idash_{ij}(\omega)\, \rho\, \Idash_{kl}^\dag(\omega)
-\frac12
\big\{\Idash_{ij}^\dag(\omega) \Idash_{kl}(\omega), \rho\big\}
\big]
\nppp&&
\hspace{-10pt}
+\ngw(\omega)
\big[
\Idash_{ij}^\dag(\omega)\, \rho\, \Idash_{kl}(\omega)
-\frac12
\big\{\Idash_{ij}(\omega) \Idash_{kl}^\dag(\omega), \rho\big\}
\big]\!
\Big\}
\label{diss2}
\end{eqnarray}
summing over $\omega>0$ and $i,j,k,l=1,2,3$.
With the aid of the formulae\refs{MTW1973}
\begin{eqnarray*}
\int\! k_i k_j \,\dd\Omega
&=&
\frac{4\pi}3\,k^2 \delta_{ij}
\end{eqnarray*}
\begin{eqnarray*}
\int\! k_i k_j k_k k_l \,\dd\Omega
&=&
\frac{4\pi}{15}\,k^2
(\delta_{ij}\delta_{kl}+\delta_{ik}\delta_{jl}+\delta_{il}\delta_{jk})
\end{eqnarray*}
we can calculate that
\begin{eqnarray}
\!\!\int\! P_{ijkl}(\kk) \,\dd\Omega
&=&
\frac{4\pi}{15}
(3\delta_{ik}\delta_{jl}+3\delta_{il}\delta_{jk}-2\delta_{ij}\delta_{kl})
\label{iPijkl} .
\end{eqnarray}
Substituting \Eq{iPijkl} into \Eq{diss2} and using the tracelessness of $\Idash_{ij}$, we obtain the following general dissipator
\begin{eqnarray}
\DD\rho
&=&
\frac{2G\omega^5}{5\hbar}
\,\times
\nppp&&
\hspace{-15pt}
\Big\{(\ngw(\omega)+1)
\big[
\Idash_{ij}(\omega)\, \rho\, \Idash_{ij}^\dag(\omega)
-\frac12
\big\{\Idash_{ij}^\dag(\omega) \Idash_{ij}(\omega), \rho\big\}
\big]
\nppp&&
\hspace{-15pt}
+\ngw(\omega)
\big[
\Idash_{ij}^\dag(\omega)\, \rho\, \Idash_{ij}(\omega)
-\frac12
\big\{\Idash_{ij}(\omega) \Idash_{ij}^\dag(\omega), \rho\big\}
\big]
\Big\}
\label{diss3}
\end{eqnarray}
of the Lindblad form for any nonrelativistic matter systems with frequency domain reduced quadrupole moments $\Idash_{ij}(\omega)$.
For particles in an isotropic harmonic potential with $\ngw(\omega)=0$, \Eq{diss3} recovers the corresponding dissipator discussed in \Ref{Oniga2017}.



\section{Gravitationally induced transitions of quantum states}
\label{sec:trans}

The theoretical framework established above is applied in this section to the quantum gravitational decoherence and radiation of a real scalar field $\phi$ with mass $m$ and the associated inverse reduced Compton wavelength $\mu = m/\hbar$, subject to an external potential $\nu(\rr)$ described by the Lagrangian density
\begin{eqnarray}
\LL
&=&
-
\frac12\,g^{\alpha\beta}\phi_{,\alpha}\phi_{,\beta}
-\Big(\frac{1}{2}+\nu\Big)\mu^2
\phi^2.
\label{LL}
\end{eqnarray}

To consider the nonrelativistic dynamics of the scalar field representing nearly Newtonian particles we assume the potential energy to be much less than the rest mass energy so that $\nu \ll 1$ and the total energy density is given approximately by
\begin{eqnarray}
T^{00}=\frac{1}{2}\big(\dot{\phi}^2+\mu^2\phi^2\big) .
\label{T00}
\end{eqnarray}

The unperturbed part of Eq.~\eqref{LL} satisfies quantum field equation
\begin{eqnarray}
\ddot\phi
=
\nabla^2\phi
-
(1+2\nu)\mu^2\phi
\label{hseqa}
\end{eqnarray}
having solutions of the convenient form
\begin{eqnarray}
\phi
=
\sqrt{\frac{\hbar}{2\omega_{n}}}\,a_{n}
\psi_n(\rr)\, e^{-i\omega_n t} + \hc
\label{hPhi}
\end{eqnarray}
with some operators $a_{n}$ and orthogonal complex functions $\psi_n(\rr)$.
Hence Eq.~\eqref{hseqa} reduces formally to the time-independent Schr\"odinger equation
\begin{eqnarray}
-\frac{\hbar^2}{2m}\nabla^2\psi_n + V\psi_n
=
E_n \psi_n
\label{hschreq}
\end{eqnarray}
where $V=m\,\nu(\rr)$ and
\begin{eqnarray}
E_n
=
\frac{1}{2m}(\hbar^2\omega_n^2-m^2)
\label{wmE}
\end{eqnarray}
represent the potential and eigenenergies respectively.

Let us consider the atom potential energy $V(r)$.
In the nonrelativistic domain, Eq.~\eqref{hPhi} becomes
\begin{eqnarray*}
\phi
&=&
\frac{\hbar}{\sqrt{2m}}\,
\big(
a_{\nn} e^{-i\omega_{\nn} t}
+
a_{\nn}^\dag  e^{i\omega_{\nn} t}
\big)
\psi_{\nn}(\rr)
\end{eqnarray*}
using the multiple indices $\nn=(n,l,m)$ and the atom wave functions
\begin{eqnarray*}
\psi_{\nn}(\rr)
&=&
R_{nl}(r)Y_l^m(\theta,\phi)
\end{eqnarray*}
with
\begin{eqnarray*}
\omega_{\nn}
&=&
\mu + E_{\nn}/\hbar
\end{eqnarray*}
arising from  the nonrelativistic limit of Eq.~\eqref{wmE}, where the corresponding ladder operators $a_{\nn}$ and $a_{\nn}^\dag$ are annihilation and creation operators respectively.
From \Eqq{qij2}{T00} we have
\begin{eqnarray}
\hspace{-12pt}
\Idash_{ij}
&=&
\frac{m}{3}\!\!\!\!\!
\sum_{\substack{{\nn,\nn'}\\{(\omega_{\nn}>\omega_{\nn'})}}}\!\!\!
\Bra{\nn'}Q_{ij}\Ket{\nn}\,
a_{\nn'}^\dag a_{\nn}\,
e^{-i(\omega_{\nn}-\omega_{\nn'}) t}
+\hc
\label{Iw}
\end{eqnarray}
where any irrelevant time-independent parts with $\omega_{\nn}=\omega_{\nn'}$ have been left out and for simplicity we have adopted the quantum mechanics styled notation
\begin{eqnarray}
\Bra{\nn'}Q_{ij}\Ket{\nn}
&=&
\int\dd^3x\,
Q_{ij}
\psi_{\nn'}(\rr)
\psi_{\nn}(\rr)
\end{eqnarray}
where Dirac's brakets are denoted by $\Bra{\cdot}$ and $\Ket{\cdot}$, as opposed to $\bra{\cdot}$ and $\ket{\cdot}$ reserved for the main quantum states of our present discussion based on quantum field theory.
Comparing \Eqq{ITT3}{Iw} we see that
\begin{eqnarray}
\Idash_{ij}(\omega_{\nn}-\omega_{\nn'})
&=&
\frac{m}{3}
\Bra{\nn'}Q_{ij}\Ket{\nn}\,
a_{\nn'}^\dag a_{\nn}
\label{Iw2a}
\ppp
\Idash_{ij}^\dag(\omega_{\nn}-\omega_{\nn'})
&=&
\frac{m}{3}
\Bra{\nn'}Q_{ij}\Ket{\nn}^*\,
a_{\nn}^\dag a_{\nn'}
\label{Iw2b}
\end{eqnarray}
for $\omega_{\nn} > \omega_{\nn'}$.


Let us apply the above general setup to a two-level atom model with states
$\ket{1} = \ket{n_1,l_1,m_1}$ and $\ket{2} = \ket{n_2,l_2,m_2}$ having respective eigen energies $E_1 = \hbar\omega_1$ and  $E_2 = \hbar\omega_2$ and the transition frequency $\omega_0=\omega_2-\omega_1>0$.
It is useful to denote these two states by
\begin{eqnarray*}
\ket{1}
=
a_{1}^\dag \ket{0}
,\quad
\ket{2}
=
a_{2}^\dag \ket{0}
\end{eqnarray*}
with the ladder operators
\begin{eqnarray*}
\sigma^{+}
&=&
\ket{2}\bra{1}
=
a_{2}^\dag a_{1}
,\quad
\sigma^{-}
=
\ket{1}\bra{2}
=
a_{1}^\dag a_{2} .
\end{eqnarray*}
Introducing the transition quadrupole moment
\begin{eqnarray}
q_{ij}
&=&
\Bra{1} Q_{ij} \Ket{2}
\label{qij}
\end{eqnarray}
we can express \Eqq{Iw2a}{Iw2b} as
\begin{eqnarray}
\Idash_{ij}(\omega_0)
&=&
\frac{m}{3}\,q_{ij}\,a_{1}^\dag a_{2}=
\frac{m}{3}\,q_{ij}\,\sigma^-
\label{Iw3a}
\ppp
\Idash_{ij}^\dag(\omega_0)
&=&
\frac{m}{3}\,q_{ij}\,a_{2}^\dag a_{1}=
\frac{m}{3}\,q_{ij}^*\,\sigma^+ .
\label{Iw3b}
\end{eqnarray}
Substituting \Eqq{Iw3a}{Iw3b} into \Eq{diss3} we obtain the dissipator for the two-level atom to be
\begin{eqnarray}
\DD\rho
&=&
\Gamma_0(\ngw(\omega_0)+1)
\big[
\sigma^- \rho\, \sigma^+
-\frac12
\big\{\sigma^+ \sigma^-, \rho\big\}
\big]
\nppp&&
+
\Gamma_0\ngw(\omega_0)
\big[
\sigma^+ \rho\, \sigma^-
-\frac12
\big\{\sigma^- \sigma^+, \rho\big\}
\big]
\label{diss4}
\end{eqnarray}
where we have introduced the gravitational spontaneous emission rate
\begin{eqnarray}
\Gamma_0=
\frac{2 G m^2 \omega_0^5 q^2}{45\hbar}
\label{gg0}
\end{eqnarray}
in terms of the coupling coefficient related to the squared modulus of the quadrupole moment of the two-level atom
\begin{eqnarray}
q^2 = q_{ij}^*q_{ij}
\label{q2}
\end{eqnarray}
subject to the following selection rules\refs{Fischer1994, Biedenhm1981},
\begin{eqnarray}
l_1+l_2 \ge 2,
\quad
\Delta l = 0,\pm2,
\quad
\Delta m = 0,\pm1,\pm2
\label{selec}
\end{eqnarray}
where $\Delta l=l_2-l_1,\Delta m=m_2-m_1$,
outside of which we have $q^2=0$.

An alternative derivation of \Eq{diss4} based on a more conventional quantum mechanical treatment is given in \apx{sec:diss}. Quantum field theory as the main approach of this work has the advantage of describing an arbitrary number of bosonic and fermionic particles and their collective behaviours more naturally and effectively.

\vspace{15pt}

\section{Gravitational wave interactions with correlated two-level atoms}
\label{sec:2level}

An $N$-particle two-level state is given by
\begin{eqnarray}
\ket{N_1,N_2}
&=&
\frac{a_{1}^{\dag N_1}}{\sqrt{N_1!}}
\frac{a_{2}^{\dag N_2}}{\sqrt{N_2!}}
\ket{0}
\label{NN}
\end{eqnarray}
with $N_1+N_2=N$. There are $N+1$ such states, which can be represented with
\begin{eqnarray}
\ket{p}=\ket{N-p+1,p-1}
\label{pp}
\end{eqnarray}
for $p=1,2,\dots N+1$, so that $\ket{1}=\ket{N,0},\ket{2}={\ket{N-1,1}},\dots\ket{N+1}=\ket{0,N}$.
It follows from
\begin{eqnarray*}
\lefteqn{\hspace{-10pt}\bra{N_1',N_2'} a_{1}^\dag a_{2} \ket{N_1,N_2}}
\nppp&=&
\sqrt{(N_1+1)N_2}\,\delta_{N_1',N_1+1}\,\delta_{N_2',N_2-1}
\ppp
\lefteqn{\hspace{-10pt}\bra{N_1',N_2'} a_{2}^\dag a_{1} \ket{N_1,N_2}}
\nppp&=&
\sqrt{N_1(N_2+1)}\,\delta_{N_1',N_1-1}\,\delta_{N_2',N_2+1}
\end{eqnarray*}
that
\begin{eqnarray*}
\bra{p'} a_{1}^\dag a_{2} \ket{p}
&=&
\sqrt{(p-1)(N-p+2)}\;\delta_{p',p-1}
\label{p1}
\ppp
\bra{p'} a_{2}^\dag a_{1} \ket{p}
&=&
\sqrt{p\,(N-p+1)}\;\delta_{p',p+1} .
\label{p2}
\end{eqnarray*}
Using the above relations, we see that \Eqq{Iw2a}{Iw2b} become
\begin{eqnarray}
\Idash_{ij}(\omega_0)
&=&
\frac{m}{3}\,q_{ij}\,\Sigma^-
\label{Iw4a}
\ppp
\Idash_{ij}^\dag(\omega_0)
&=&
\frac{m}{3}\,q_{ij}^*\Sigma^+
\label{Iw4b}
\end{eqnarray}
using the $(N+1)$-dimensional ladder operators
\begin{eqnarray}
\Sigma^+_{p' p}
=
\Sigma^-_{p\, p'}
&=&
\sqrt{p\,(N-p+1)}\;\delta_{p',p+1} .
\end{eqnarray}
Substituting \Eqq{Iw4a}{Iw4b} into \Eq{diss3}, we obtain the dissipator for the correlated $N$ two-level atoms
\begin{eqnarray}
\DD\rho
&=&
\Gamma_0(\ngw(\omega_0)+1)
\big[
\Sigma^- \rho\, \Sigma^+
-\frac12
\big\{\Sigma^+ \Sigma^-, \rho\big\}
\big]
\nppp&&
+
\Gamma_0\ngw(\omega_0)
\big[
\Sigma^+ \rho\, \Sigma^-
-\frac12
\big\{\Sigma^- \Sigma^+, \rho\big\}
\big].
\label{diss5}
\end{eqnarray}
For $N=1$, we have the reduction $\Sigma^\pm = \sigma^\pm$. However, in general, we see that
\begin{eqnarray}
\Sigma^\pm_{p\pm1 p}\approx \frac{N}{4},
\quad\hbox{ for }
p \approx \frac{N}{2}
\end{eqnarray}
with a large $N$, providing an $N^2$ amplification for \Eq{diss5} compared with \Eq{diss4}.

Consider a diagonal density matrix
\begin{eqnarray}
\rho_{p q}
&=&
\rho_p\delta_{p,q}
\label{dia}
\end{eqnarray}
with the normalization
\begin{eqnarray}
\sum_{p=1}^{N+1}
\rho_p
&=&
1
\label{norm}
\end{eqnarray}
from which we can calculate the following:
\begin{eqnarray*}
(\Sigma^+\Sigma^-\rho)_{p q}
&=&
(p-1)\,(N-p+2)\,\rho_p\,\delta_{p,q}
\nppp
(\Sigma^-\Sigma^+\rho)_{p q}
&=&
p\,(N-p+1)\,\rho_{p}\,\delta_{p,q}
\nppp
(\Sigma^-\rho\,\Sigma^+)_{p q}
&=&
p\,(N-p+1)\,\rho_{p+1}\,\delta_{p,q}
\nppp
(\Sigma^+\rho\,\Sigma^-)_{p q}
&=&
(p-1)\,(N-p+2)\,\rho_{p-1}\,\delta_{p,q}
\nppp
(\rho\,\Sigma^-\Sigma^+)_{p q}
&=&
p\,(N-p+1)\,\rho_{p}\,\delta_{p,q}
\nppp
(\rho\,\Sigma^+\Sigma^-)_{p q}
&=&
(p-1)\,(N-p+2)\,\rho_p\,\delta_{p,q}.
\end{eqnarray*}
Substituting the above relations into \Eq{diss5}, we have
\begin{eqnarray}
(\DD\rho)_{p q}
&=&
\Gamma_0\,D_p\,\delta_{p,q}
\label{diss6}
\end{eqnarray}
where
\begin{eqnarray}
D_p
&=&
(1+\ngw(\omega_0))
p\,(N-p+1)\,\rho_{p+1}
\nppp&&
-(1+\ngw(\omega_0))
(p-1)\,(N-p+2)\,\rho_p
\nppp&&
+\ngw(\omega_0)
(p-1)\,(N-p+2)\,\rho_{p-1}
\nppp&&
-\ngw(\omega_0)
p\,(N-p+1)\,\rho_{p} .
\label{fp}
\end{eqnarray}
The trace of \Eq{diss6} vanishes
\begin{eqnarray}
\tr(\DD\rho)
&=&
\sum_{p=1}^{N+1}
(\DD\rho)_{p p}
=0
\label{trdiss6}
\end{eqnarray}
as required for the consistency of \Eq{norm} under time evolution.
Denoting by $\rho(s)=N\rho_p$, $D(s)=N D_p$ using ${s=(p-1)/N}$ and ${\epsilon=1/N}$ with
${0 \le s \le 1}$,
we see that \Eq{fp} becomes
\begin{eqnarray*}
\lefteqn{D(s+\epsilon)}
\nppp
&=&
\big[
(s+\epsilon)\,(1-s)\,\rho(s+2\epsilon)
\nppp&&
-
s\,(1-(s-\epsilon))\,\rho(s+\epsilon)
\big]/\epsilon^2
\nppp&&
+
\ngw(\omega_0)
\big[
(s+\epsilon)\,(1-s)\,(\rho(s+2\epsilon)-\rho(s+\epsilon))
\nppp&&
-
s\,(1-(s-\epsilon))\,(\rho(s+\epsilon)-\rho(s))
\big]/\epsilon^2 .
\end{eqnarray*}
For a large particle number $N \gg 1$ and hence $\epsilon \ll 1$, the above tends to the continuous limit
\begin{eqnarray}
D(s)
=
\big\{
s(1-s)
\big[N\rho(s)
+
\ngw(\omega_0)
\rho'(s)
\big]
\big\}'
\label{fs}
\end{eqnarray}
where $({}') = \partial_s$. Then the discrete dissipator\eq{diss5} becomes the following continuous ``diffuser,''
\begin{eqnarray}
\DD\rho(s)
&=&
\Gamma_0
\big\{
s(1-s)
\big[N\rho(s)
+
\ngw(\omega_0)
\rho'(s)
\big]
\big\}'
\label{diss7}
\end{eqnarray}
with possible collective amplification by a factor up to $N^2$ represented by $({}') \sim N$ and $({}'') \sim N^2$.
In terms of \Eq{fs}, the traceless condition\eq{trdiss6} then becomes
\begin{eqnarray}
\int_0^1 \! D(s) \,\dd s
&=&
0 .
\label{trdiss7}
\end{eqnarray}
For a stationary finite solution with $\DD\rho=0$ and $\ngw(\omega_0) > 0$, \Eq{diss7} reduces to
\begin{eqnarray}
\ngw(\omega_0)
\rho'(s)
=
-N\rho(s)
\label{d7z2}
\end{eqnarray}
yielding
\begin{eqnarray}
\rho(s)
=
\frac{e^{-s/w}}{w(1-e^{-1/w})}
\label{rs}
\end{eqnarray}
where $w=\ngw(\omega_0)/N$.
See \Fig{fig3} for illustrative numerical examples of dynamical relaxations into the stationary states described above.

\vspace{10pt}

\section{Optimal gravitational wave interactions with correlated atoms}
\label{sec:gw}

In this section, we would like to relate the above theoretical description to physically relevant values.
To this end, let us first relate the distribution function $\ngw(\omega)$ to the spectral function $\Omega_\gw(f)$ for gravitational waves.

The gravitational wave energy density
\begin{eqnarray*}
\rho_\gw
&=&
\frac1{32\pi G}
\langle
\dot{h}_{ij}\dot{h}_{ij}
\rangle
\end{eqnarray*}
is related to $\ngw(k)$ by\refs{Schafer1981}:
\begin{eqnarray*}
\rho_\gw
&=&
\frac{\hbar}{\pi^2}
\int_0^\infty
\dd \omega\, \omega^3 \ngw(\omega) .
\label{ugw2}
\end{eqnarray*}
On the other hand, the energy density spectral function $\rho_\gw(\omega)$ of the gravitational waves is also given by
\begin{eqnarray*}
\dd \rho_\gw
&=&
\rho_\gw(\omega)\,\dd\omega.
\label{uw}
\end{eqnarray*}
Comparing  the above two relations, we see that
\begin{eqnarray*}
\rho_\gw(\omega)
&=&
\frac{\hbar}{\pi^2}\, \omega^3 \ngw(\omega) .
\end{eqnarray*}
Using the wave frequency $f=\omega/\,2\pi$, Planck constant $h=2\pi\hbar$,
and
$
\dd \rho_\gw
=
\rho_\gw(f)\,\dd f
$,
we can write the above as
\begin{eqnarray*}
\rho_\gw(f)
&=&
8\pi h\, f^3 \ngw(\omega) .
\end{eqnarray*}
Then, by using the critical energy density $\rhoc$ in cosmology, we see that the dimensionless spectral function $\Omega_\gw(f)$ for gravitational waves is given by
\begin{eqnarray}
\Omega_\gw(f)
&=&
\frac1{\rhoc}
\frac{\dd \rho_\gw}{\dd \ln f}
=
\frac{8\pi h}{\rhoc}\, f^4 \ngw(\omega) .
\label{Omf}
\end{eqnarray}
To recover $\ngw(\omega)$ from given $\Omega_\gw(f)$, we can invert \Eq{Omf} to get
\begin{eqnarray}
\ngw(\omega)
&=&
\frac{\rhoc}{8\pi h}
\frac{\Omega_\gw(f)}{f^4}
\label{Omfa}
\end{eqnarray}
which is the sought relation between the distribution function $\ngw(\omega)$ and spectral function $\Omega_\gw(f)$.

Let us now consider an estimate of the single atom transition rate $\Gamma_0$ with states
$\ket{1} = \ket{n_1,l_1,m_1}$ and $\ket{2} = \ket{n_2,l_2,m_2}$
where
$n_2 \ge n_1$

In terms of the Bohr radius $a_0$, we can make the following approximations
\begin{eqnarray*}
\Bra{R_{n_1,l_1}}r^2\Ket{R_{n_1,l_1}}
&\approx&
n_1^2\, a_0^2
\ppp
\Bra{R_{n_2,l_2}}r^2\Ket{R_{n_2,l_2}}
&\approx&
n_2^2\, a_0^2
\end{eqnarray*}
and hence\refs{Fischer1994}
\begin{eqnarray}
\Bra{R_{n_1,l_1}}r^2\Ket{R_{n_2,l_2}}^2
&\approx&
n_1^2 n_2^2\, a_0^4.
\label{a4}
\end{eqnarray}
Substituting \Eq{a4} into \Eq{q2} and using the related 3-$j$ symbols described in \apx{sec:3j} we can therefore approximate
\begin{eqnarray}
q^2
\approx
n_1^2 n_2^2\, a_0^4
\end{eqnarray}
subject to the selection rules\eq{selec}.
Then the gravitational spontaneous emission rate for a single atom given by \Eq{gg0}, with the speed of light $c$ restored,  reads
\begin{eqnarray}
\Gamma_0 =
\frac{2\,G m^2 n_1^2 n_2^2\, a_0^4\, \omega_0^5}{45\,\hbar\, c^5}
\label{gg0a}
\end{eqnarray}
also subject to the selection rules\eq{selec}.

To take advantage of an $N^2$ amplification factor for the transition rate discussed in Sec.~\ref{sec:2level}, we will take as initial state a spikelike profile specified by
\begin{eqnarray}
\rho_p
=
\delta_{p,p_0}
\label{drp}
\end{eqnarray}
with $1\le p_0 \le N+1$; then, \Eq{fp} becomes
\begin{eqnarray}
D_p
&=&
(1+\ngw(\omega_0))
p\,(N-p+1)\,\delta_{p+1,p_0}
\nppp&&
-(1+\ngw(\omega_0))
(p-1)\,(N-p+2)\,\delta_{p,p_0}
\nppp&&
+\ngw(\omega_0)
(p-1)\,(N-p+2)\,\delta_{p-1,p_0}
\nppp&&
-\ngw(\omega_0)
p\,(N-p+1)\,\delta_{p,p_0} .
\label{fp2}
\end{eqnarray}
The transition rate of the initial state follows as
\begin{eqnarray}
D_{p_0}
&=&
-(1+\ngw(\omega_0))
(p_0-1)\,(N-p_0+2)
\nppp&&
-\ngw(\omega_0)
p_0\,(N-p_0+1)
\label{fp2a}
\end{eqnarray}
whereas the transition rate of the state above the initial state follows as
\begin{eqnarray}
D_{p_0+1}
&=&
\ngw(\omega_0)
p_0\,(N-p_0+1) .
\label{fp2b}
\end{eqnarray}
For a large $N$, to maximize the absolute values of \Eqq{fp2a}{fp2b}, we choose $p_0\approx N/2$, and then we have
\begin{eqnarray}
D_{p_0}
&=&
-\frac{N^2}{4}(1+2\ngw(\omega_0))
\label{fp2a0}
\end{eqnarray}
and
\begin{eqnarray}
D_{p_0+1}
&=&
\frac{N^2}{4}
\ngw(\omega_0)
\label{fp2b0}
\end{eqnarray}
to the leading order in $N$.
The short-time evolutions of such ``favourable states'' under gravitational waves and spacetime fluctuations are illustrated in \Figg{fig3}{fig4}.

Using \Eqqqq{diss6}{gg0a}{fp2a0}{fp2b0}, it is convenient to introduce the effective scale transition rates
\begin{eqnarray}
\Gamma_\gw
=
\frac{N^2}{4}\,
\Gamma_0
\ngw(\omega_0)
\label{gggw}
\ppp
\Gamma_\vac
=
\frac{N^2}{4}\,
\Gamma_0\nvac(\omega_0)
\label{ggvac}
\end{eqnarray}
and the associated relaxation times
\begin{eqnarray}
\tau_\gw
=
\frac1{\Gamma_\gw}
,\quad
\tau_\vac
=
\frac1{\Gamma_\vac}
\label{tt}
\end{eqnarray}
due respectively to stochastic gravitational waves and vacuum fluctuations, in terms of the constant vacuum distribution function
\begin{eqnarray}
\nvac(\omega)=1 .
\label{nvac}
\end{eqnarray}

\Eqq{Omf}{nvac} allow us to give an analogous spectral function for vacuum fluctuations as a quartic function of frequency to be
\begin{eqnarray}
\Omega_\vac(f)
&=&
\frac{8\pi h f^4}{c^3 \rhoc}
\label{Omvac}
\end{eqnarray}
with the speed of light $c$ restored.

\begin{figure}[!ht]
\vspace{40pt}
\includegraphics[width=1\linewidth]{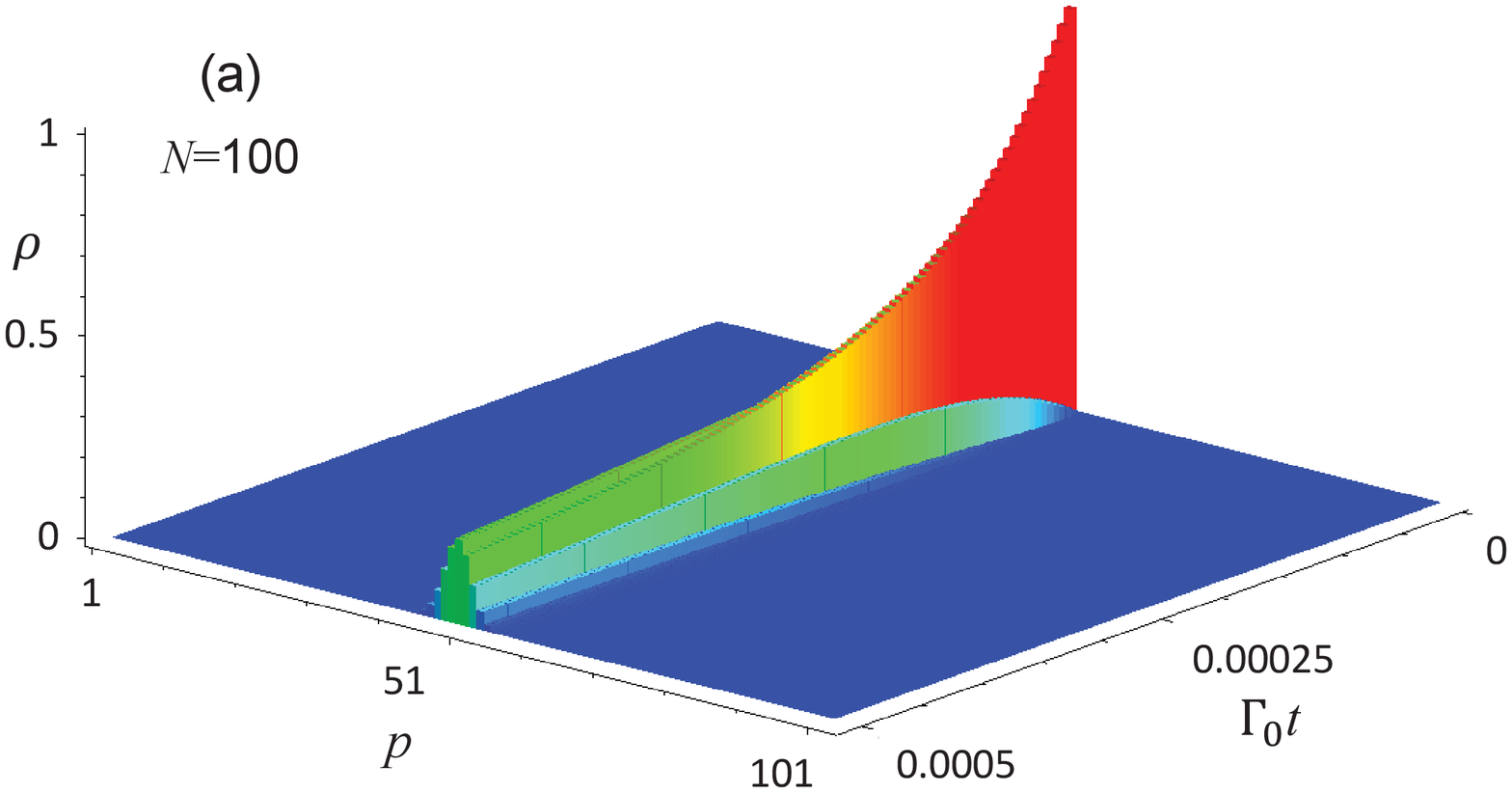}\\
\vspace{40pt}
\includegraphics[width=1\linewidth]{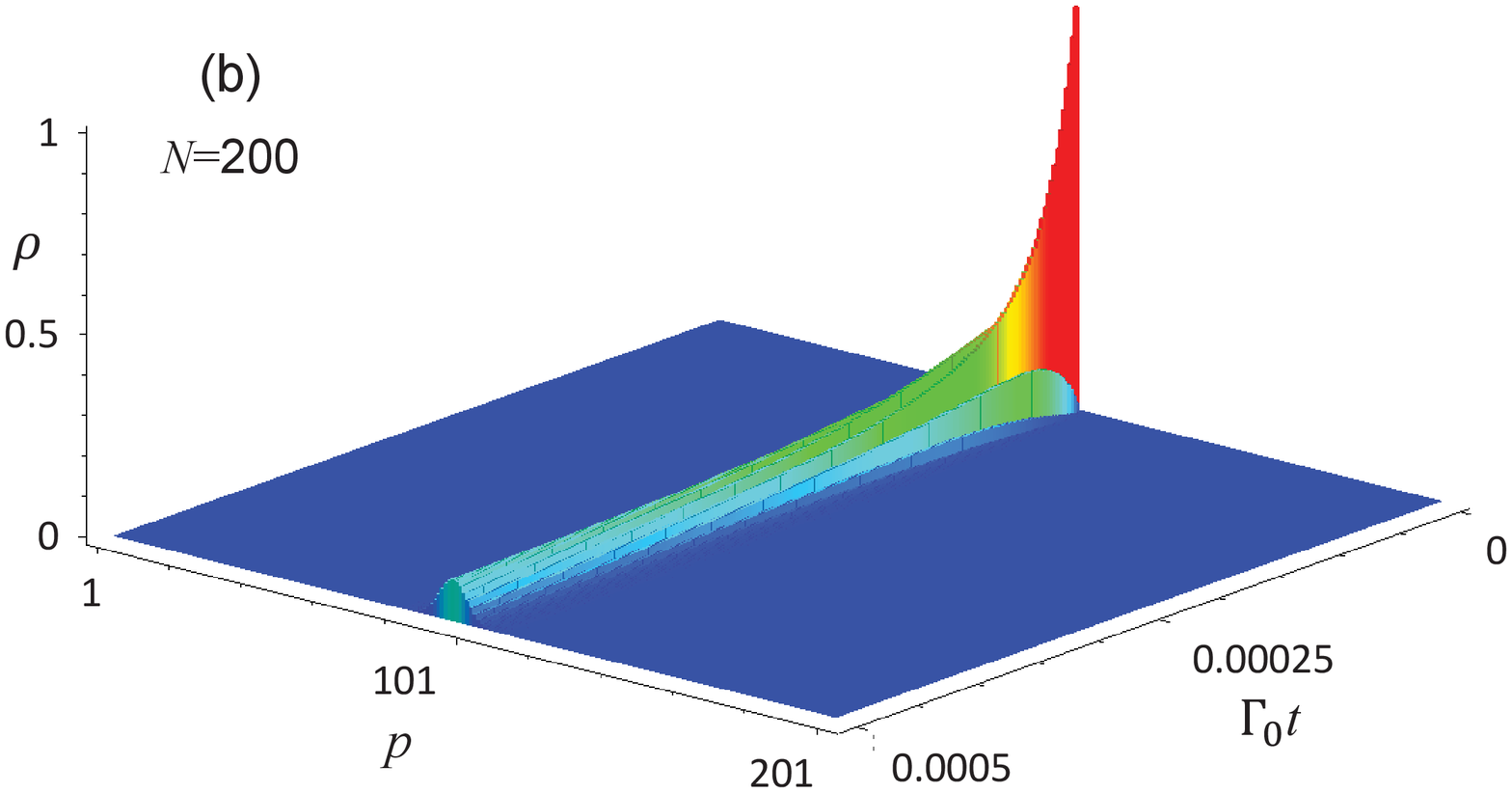}
\caption{Simulations of short-time collective transitions due to vacuum fluctuations of spacetime and additional stochastic gravitational waves with a chosen $\ngw(\omega_0)=1$.
In plot (a) with $N=100$ atoms, the initial state corresponds to the spike of height 1 at $(p, t)=(50,0)$. While this state decays similarly to the descriptions in \Fig{fig4}, its adjacent higher state with $p=51$ is excited with an initial time scale of $\tau_\gw=4/({N^2\Gamma_0}\ngw(\omega_0))=0.0004/\Gamma_0$ according to \Eqq{gggw}{tt}.
In plot (b) with $N=200$ atoms, the initial state corresponds to the spike of height 1 at $(p, t)=(100,0)$. For a doubled $N$,
its adjacent higher state with $p=101$ is excited with a quarter of the preceding initial time scale with $\tau_\gw=4/({N^2\Gamma_0}\ngw(\omega_0))=0.0001/\Gamma_0$.
}
\label{fig4}
\vspace{40pt}
\end{figure}

If we envisage an off-resonant high-$Q$ cavity of volume $L^3$, then the number of contained atoms $N$ is limited by
\begin{eqnarray}
N
&\lesssim&
\frac{L^3}{4\pi(n_2^2 a_0)^3/3}.
\label{NL}
\end{eqnarray}
Substituting \Eqqqq{gg0a}{gggw}{tt}{NL} into \Eq{Omf}, we have
\begin{eqnarray}
\Omega(f_0)
&\gtrsim&
\frac{160\,\hbar^2 c^2 a_0^2 \,n_2^{10}}
{G L^6 \rhoc m^2\,\tau\, \omega_0\,n_1^2}
\label{Omf2}
\end{eqnarray}
which applies to $\Omega=\Omega_\gw$ and $\Omega_\vac$ for given $\tau=\tau_\gw$ and $\tau_\vac$ for the stochastic gravitational waves and vacuum spacetime fluctuations respectively.
Considering the transition frequency given by
\begin{eqnarray}
\omega_0
&=&
\Big(\frac{1}{n_1^2}-\frac{1}{n_2^2}+\epsilon\Big)\frac{\ER}{\hbar}
\label{f0ER}
\end{eqnarray}
where $\epsilon$ indicates an energy shift due to, e.g., quantum defects or Lamb/Zeeman shifts, it is clear that the lower bound of the spectral function \Eq{Omf2} is minimized for $f_0 \lesssim$ THz with $\epsilon \ll 1$ by {\it small} principal quantum numbers $n_2=n$ which is as close to 1 as possible, with $n_1$ equal to or as close to $n$ as possible.

Subject to further experimental constraints to be investigated further, \Eq{Omf2} provides theoretical lower bounds on their spectral functions for detection, as preluded in \Sec{sec:intr} with \Fig{fig1}, where specific numerical examples using an ensemble of heliumlike atoms are studied with $n_1=n_2=n=2$, $l_1=l_2=1$, $-m_1=m_2=1$, satisfying the selection rules\eq{selec} with $l_1+l_2=2, \Delta l=0 , \Delta m=2$, and ${0 < \epsilon \lesssim 0.003}$, which could be induced by a Zeeman-type shift.

\vspace{15pt}

\section{Conclusion and discussion}
\label{sec:conl}

Recent developments in gravitational and quantum physics have not only enriched applications with emerging prospects but also escalated the quest for a unified theory of quantum gravity that in turn has important bearing on the structure and evolution of the Universe. The growing availabilities for quantum matter exemplified by ultracold atoms with higher excitations, more correlations, better controls, and in larger quantities promise valuable tools in this scientific endeavour.

In this work, we have carried out new theoretical analysis describing how a large number of correlated atoms could interact and sense gravitational waves of fundamental and cosmological origins to go beyond the physical capacities of conventional detectors.
We have reported theoretical results within generic physical constraints, but not fully taken onboard experimental implementations, which are a subject of future work.

Nonetheless, apart from extending the conceptual framework, building up theoretical tools and guiding possible viable detection of gravitational waves outside existing windows including zero-point fluctuations, our findings can already be used to rule out potential claims of effects that can be mapped to below our theoretical lower bounds on measurable gravitational wave spectral function values using quantum techniques, as noted in the caption of \Fig{fig1}, unless new inputs in addition to considerations adopted here are supplied.

In going forward, it would also be worth exploring the effects of the interaction Hamiltonian $H_\intt$ in \Eq{maseqn}, which we neglect at present for simplicity, to accommodate atom-atom interactions, external control of atoms, and interactions with regular and nonstochastic gravitational waves for their quantum sensing based our framework.

\vspace{15pt}

\acknowledgments

The authors are grateful for financial support to the National Council for Science and Technology (CONACyT) (D.Q.), the Carnegie Trust for the Universities of Scotland (T.O.), and the Cruickshank Trust and EPSRC GG-Top Project (C.W.).

\vspace{15pt}

\appendix

\section{Alternative derivation of the quantum dissipator of the two-level atom}
\label{sec:diss}

The quantum mechanical interaction Hamiltonian between a mass $m$ and gravitational waves\refs{Fischer1994, Parker1980} is given by
\begin{eqnarray}
H_I
&=&
\frac12 m R_{0i0j} r_i r_j .
\label{GWI}
\end{eqnarray}
Using the weak gravitational wave relation\refs{Flanagan2005}
\begin{eqnarray*}
R_{0i0j} = -\frac12 \ddot{h}^\TT_{ij}
\end{eqnarray*}
we see that Eq. \eqref{GWI} becomes
\begin{eqnarray}
H_I
&=&
-\frac1{12} m Q_{ij} \,\ddot{h}^\TT_{ij}
\label{GWI2}
\end{eqnarray}
where $Q_{ij}$ is the quadrupole moment tensor given by \Eq{Qij}.

It is useful to write \Eq{GWI2} in the standard form of the interaction Hamiltonian for open quantum systems\refs{Breuer2002}
\begin{eqnarray}
H_I
&=&
\sum_\alpha
A_\alpha B_\alpha
\label{GWI3}
\end{eqnarray}
with the following identifications:
\begin{eqnarray}
\alpha&\to&(i,j)
\label{idal}
\ppp
A_\alpha&\to& A_{ij} = m  Q_{ij}
\label{idA}
\ppp
B_\alpha&\to& B_{ij} = -\frac1{12} \ddot{h}^\TT_{ij}
\label{idB}
\end{eqnarray}
yielding in our two-level case the following Lindblad operators
\begin{eqnarray}
A_{ij}(\omega)
&=&
\Ket{1}\Bra{1} A_{ij} \Ket{2}\Bra{2}
=
m\, q_{ij}\, \sigma^{-}
\label{Ap}
\ppp
A_{ij}(-\omega)
&=&
\Ket{2}\Bra{2} A_{ij} \Ket{1}\Bra{1}
=
m\, q_{ij}^*\, \sigma^{+}
\label{Am}
\end{eqnarray}
using \Eq{qij} and
\begin{eqnarray*}
\sigma^{+}
&=&
\Ket{2}\Bra{1}
,\quad
\sigma^{-}
=
\Ket{1}\Bra{2} .
\end{eqnarray*}
Clearly, $A_{ij}(\omega)$ satisfies
\begin{eqnarray}
A_{ij}^\dag(\omega)=A_{ij}(-\omega),\quad A_{kk}(\omega)=0
\label{Aw}
\end{eqnarray}
for $\omega\to\pm\omega$.

Using the rotating wave approximation, we have the following dissipator
\begin{eqnarray}
\DD\rho
&=&
\sum_{\omega\to\pm\omega}  \sum_{\alpha,\alpha'}
\Gamma_{\alpha,\alpha'}(\omega)\times
\nppp&&
\hspace{-15pt}
\left [A_{\alpha}(\omega) \rho A_{\alpha'}^\dag (\omega) - A_{\alpha'}^\dag (\omega) A_{\alpha}(\omega) \rho \right ] + \hc
\label{master2}
\end{eqnarray}
with the spectral correlation tensor
\begin{eqnarray}
\Gamma_{\alpha,\alpha'}(\omega)
&=&
\frac{1}{\hbar^2} \int_0^{\infty} \dd s\,
e^{i \omega s} \langle{B_{\alpha}^\dag(s) B_{\alpha'}(0)} \rangle
\label{Gaa}
\end{eqnarray}
using a Gaussian stationary bath of gravitational waves.
This expression can be evaluated by using \Eq{idB} with the field expansion for $h_{ij}^\TT$.
To evaluate \Eq{Gaa} consider the  wave expansion for the gravitons
\begin{eqnarray}
{h}^\TT_{ij}
=
\int\dd^3 k
\sqrt{\frac{G\hbar}{\pi^2 k}}\,
e_{ij}^\lambda(\kk)\,
a_{\kk}^\lambda\,
e^{ikx}
+
\hc
\label{gwexa}
\end{eqnarray}
with the relations
\begin{eqnarray*}
\big[
a_{\kk}^\lambda, a_{\kk'}^{\lambda'\dag}
\big]
&=&
\delta_{\lambda\lambda'}\delta(\kk, \kk')
\ppp
\big[
a_{\kk}^{\lambda\dag}, a_{\kk'}^{\lambda'\dag}
\big]
&=&
\big[
a_{\kk}^\lambda, a_{\kk'}^{\lambda'}
\big]
=0 .
\end{eqnarray*}

To proceed, consider the discrete momentum $\kb$ of the atom in a box of volume $V$ expressed in terms of the Cartesian wave numbers $n_i = 1, 2, \cdots$ as follows:
\begin{eqnarray*}
\kb
&=&
\left(\frac{2 \pi n_1}{V^{1/3}},
\frac{2 \pi n_2}{V^{1/3}},
\frac{2 \pi n_3}{V^{1/3}}\right)
\end{eqnarray*}
In order to write the field in this box, we use
\begin{eqnarray}
\int \dd^3 k \leftrightarrow \frac{(2 \pi)^3}{V} \sum_{\kb}
\label{ds}
\end{eqnarray}
with the modes inside the box normalized such that
\begin{eqnarray*}
\int_V \dd ^3 x \; e^{i \kb \rr} e^{- i \kb' \rr} = V \delta_{\kb \kb'} .
\end{eqnarray*}
Along with the following representation for the delta function in a finite volume,
\begin{eqnarray*}
\delta(\rr - \rr') = \frac{1}{V} \sum_{\kb} e^{i \kb (\rr - \rr')}
\end{eqnarray*}
The Dirac delta function maps to the Kronecker delta as follows:
\begin{eqnarray*}
\delta(\kk-\kk')
&\leftrightarrow&
\frac{V}{(2 \pi)^3}\,\delta_{\kb \kb'} .
\end{eqnarray*}
Ladder operators for free fields are related by
\begin{eqnarray*}
a_{\kk}
&\leftrightarrow&
\sqrt{\frac{V}{(2 \pi)^{3}}}\; a_{\kb}
\\
a_{\kk}^\dagger
&\leftrightarrow&
\sqrt{\frac{V}{(2 \pi)^{3}}}\; a_{\kb}^\dag
\end{eqnarray*}
so that
\begin{eqnarray*}
[a_{\kk}, a_{\kk'}^\dagger]
&=&
\delta(\kk-\kk')
\nppp
[a_{\kb}, a_{\kb}^\dagger]
&=&
\delta_{\kb \kb'} .
\end{eqnarray*}
Then, the discrete version of \eqref{gwexa} takes the form
\begin{eqnarray}
{h}^\TT_{ij}
&=&
\sum_{\kb}\sqrt{\frac{8\pi G \hbar}{V k}}\,
e_{ij}^\lambda(\kb)\,
a_{\kb}^\lambda\,
e^{ikx}
+
\hc
\label{gwexb}
\end{eqnarray}
For an equilibrium environment\refs{Breuer2002}, we have
\begin{eqnarray*}
\langle
a_{\kb}^{\lambda\dag} a_{\kb'}^{\lambda'}
\rangle
&=&
\delta_{\lambda\lambda'}\delta_{\kb, \kb'}\,\ngw(k)
\ppp
\langle
a_{\kb}^\lambda a_{\kb'}^{\lambda'\dag}
\rangle
&=&
\delta_{\lambda\lambda'}\delta_{\kb, \kb'}(1+\ngw(k))
\ppp
\langle
a_{\kb}^{\lambda\dag} a_{\kb'}^{\lambda'\dag}
\rangle
&=&
\langle
a_{\kb}^\lambda a_{\kb'}^{\lambda'}
\rangle
=0
\end{eqnarray*}
for some distribution function $\ngw(k)$ with the quantum ensemble average $\av{\cdot}$.
Then using Eqs.~\eqref{idB}, \eqref{gwexb}, and the above, Eq. \eqref{Gaa} becomes
\begin{eqnarray*}
\Gamma_{ij,kl}(\omega)
&=&
\frac{\pi G k^3}{9\hbar V}\,
\sum_{\kb}
P_{ijkl}(\kb)\times
\nppp&&\hspace{-30pt}
\int_0^{\infty} \dd s\,[
(1+\ngw(k))\,e^{-i(k-\omega)s}
+
\ngw(k)\,e^{i(k+\omega)s}
]
\end{eqnarray*}
Then by using \Eq{ds}, the above returns to the continuous momentum case as follows
\begin{eqnarray*}
\Gamma_{ij,kl}(\omega)
&=&
\frac{G}{72\pi^2\hbar}
\int_0^\infty \dd k \,k^5
\int \dd \Omega \,
P_{ijkl}(\kk)\times
\nppp&&\hspace{-35pt}
\int_0^{\infty} \dd s\,[
(1+\ngw(k))\,e^{-i(k-\omega)s}
+
\ngw(k)\,e^{i(k+\omega)s}
] .
\end{eqnarray*}
Substituting the relation\eq{iPijkl} into the above, we have
\begin{eqnarray*}
\Gamma_{ij,kl}(\omega)
&=&
\frac{G(3\delta_{ik}\delta_{jl}+3\delta_{il}\delta_{jk}-2\delta_{ij}\delta_{kl})}
{270\pi\hbar}
\int_0^\infty \!\dd k \,k^5
\times
\nppp&&\hspace{-30pt}
\int_0^{\infty} \!\!\dd s [
(1+\ngw(k))\,e^{i(\omega-k)s}
+
\ngw(k)\,e^{i(\omega+k)s}
]
\end{eqnarray*}
Using the the Sokhotski-Plemelj theorem\refs{Breuer2002}
\begin{equation}
\int_0^{\infty} \dd s\, e^{-i \epsilon s} = \pi \delta(\epsilon) - i \text{P} \frac{1}{\epsilon}
\label{fma}
\end{equation}
where P denotes the Cauchy principal value, the above becomes
\begin{eqnarray*}
\Gamma_{ij,kl}(\omega)
&=&
\frac{G(3\delta_{ik}\delta_{jl}+3\delta_{il}\delta_{jk}-2\delta_{ij}\delta_{kl})}
{270\pi\hbar}\times
\nppp&&
\hspace{-30pt}
\int_0^\infty \dd k \,k^5
\Big[
(1+\ngw(k))
\Big(
\pi \delta(\omega-k) + i\text{P}\frac{1}{\omega - k }
\Big)
\nppp&&
\hspace{-30pt}
+
\ngw(k)
\Big(
\pi \delta(\omega+k) + i\text{P}\frac{1}{\omega + k }
\Big)
\Big] .
\end{eqnarray*}
Neglecting P terms as they do not contribute to quantum decoherence and dissipation, we see that the above reduces to
\begin{eqnarray*}
\Gamma_{ij,kl}(\omega)
&=&
\frac{G(3\delta_{ik}\delta_{jl}+3\delta_{il}\delta_{jk}-2\delta_{ij}\delta_{kl})}
{270\hbar}\int_0^\infty \!\dd k
\times
\nppp&&
\hspace{-20pt}
k^5\Big[
(1+\ngw(k))\delta(\omega-k)
+
\ngw(k)\delta(\omega+k)
\Big]
\end{eqnarray*}
yielding
\begin{eqnarray*}
\Gamma_{ij,kl}(\omega)
&=&
\frac{G\omega^5}{270\hbar}\times
\nppp&&
(3\delta_{ik}\delta_{jl}+3\delta_{il}\delta_{jk}-2\delta_{ij}\delta_{kl})
(\ngw(\omega)+1)
\nppp
\Gamma_{ij,kl}(-\omega)
&=&
\frac{G\omega^5}{270\hbar}
(3\delta_{ik}\delta_{jl}+3\delta_{il}\delta_{jk}-2\delta_{ij}\delta_{kl})
\ngw(\omega) .
\end{eqnarray*}
Substituting these two relations into the dissipator \eqref{master2} and using \eqref{Aw}, we have
\begin{eqnarray*}
\DD\rho
&=&
\frac{G\omega^5}{90\hbar}
\sum_{ij,kl}
(\delta_{ik}\delta_{jl}+\delta_{il}\delta_{jk})
\times
\nppp&&
\Big\{
(\ngw(\omega)+1)
\big[
A_{ij}(\omega) \rho A_{kl}^\dag (\omega)
-
A_{kl}^\dag (\omega) A_{ij}(\omega) \rho
\nppp&&
+
A_{kl}(\omega) \rho A_{ij}^\dag (\omega)
-
\rho A_{ij}^\dag(\omega) A_{kl} (\omega)
\big]
\nppp&&
+
\ngw(\omega)
\big[
A_{ij}^\dag(\omega) \rho A_{kl} (\omega)
-
A_{kl}(\omega) A_{ij}^\dag(\omega) \rho
\nppp&&
+
A_{kl}^\dag (\omega) \rho A_{ij}(\omega)
-
\rho A_{ij}(\omega) A_{kl}^\dag(\omega)
\big]
\Big\}
\end{eqnarray*}
which yields the following dissipator of the Lindblad form:
\begin{eqnarray*}
\DD\rho
&=&
\frac{2\,G\omega^5}{45\hbar}
\sum_{ij}
\Big\{
(\ngw(\omega)+1)\times
\\&&\hspace{-10pt}
\big [
A_{ij}(\omega) \rho A_{ij}^\dag (\omega)
-
\frac12 \big\{A_{ij}^\dag (\omega) A_{ij}(\omega),\, \rho\big\}
\big ]
\nppp&&\hspace{-10pt}
+
\ngw(\omega)
\big[
A_{ij}^\dag(\omega) \rho A_{ij} (\omega)
-
\frac12 \big\{A_{ij}(\omega) A_{ij}^\dag(\omega),\, \rho\big\}
\big]
\Big\} .
\end{eqnarray*}
Finally, by using \Eqq{Ap}{Am}, we can further simplify the above to the same form as \Eq{diss4}.

\vspace{15pt}

\section{Evaluation of the transition quadrupole moment}
\label{sec:3j}

To evaluate $q_{ij}$ in \Eq{qij}, let us consider the components of $Q_{ij}$ given by \Eq{Qij}, noting that $Q_{ij}$ has five independent components for being traceless.
For $l=2$, we can express $Q_{ij}$ in terms of the spherical harmonics $Y_{l=2}^{m}(\theta,\phi)$ as follows:
\begin{eqnarray}
Q_{11}
&=&
\sqrt{\frac{6\pi}5}\, r^2
(Y_2^{-2}+Y_2^{2})
-
\sqrt{\frac{4\pi}{5}}\, r^2\,
Y_2^{0}
\label{Q11}
\ppp
Q_{22}
&=&
-\sqrt{\frac{6\pi}5}\, r^2
(Y_2^{-2}+Y_2^{2})
-
\sqrt{\frac{4\pi}{5}}\, r^2\,
Y_2^{0}
\ppp
Q_{33}
&=&
\sqrt{\frac{16\pi}{5}}\, r^2\,
Y_2^{0}
\ppp
Q_{12}
&=&
i\sqrt{\frac{6\pi}5}\, r^2
(Y_2^{-2}-Y_2^{2})
\ppp
Q_{13}
&=&
\sqrt{\frac{6\pi}5}\, r^2
(Y_2^{-1}-Y_2^{1})
\ppp
Q_{23}
&=&
i\sqrt{\frac{6\pi}5}\, r^2
(Y_2^{-1}+Y_2^{1})
\label{Q23} .
\end{eqnarray}
Denoting by
\begin{eqnarray*}
\Bra{R_{n_1,l_1}}f(r)\Ket{R_{n_2,l_2}}
=
\int_{0}^\infty\!\dd r\,r^2 f(r) R_{n_1,l_1}(r) R_{n_2,l_2}(r)
\end{eqnarray*}
we can evaluate the matrix elements of the above $Q_{ij}$ in terms of Wigner's 3-$j$ symbols through the relation
\begin{eqnarray}
\lefteqn{\Bra{n_1,l_1,m_1} r^2 Y_{l=2}^m(\theta,\phi) \Ket{n_2,l_2,m_2}}
\nppp
&=&
\Bra{R_{n_1,l_1}}r^2\Ket{R_{n_2,l_2}}
(-1)^{m_1}
\sqrt{\frac{(2l_1+1)(2l+1)(2l_2+1)}{4\pi}}
\nppp&&\times\;
\begin{pmatrix}
  l_1 \;& l \;& l_2 \\
  0   \;& 0 \;& 0
\end{pmatrix}
\begin{pmatrix}
  l_1 \;& l \;& l_2 \\
  -m_1\;& m \;& m_2
\end{pmatrix}
\label{1Y2}
\end{eqnarray}
with $l=2$.

\end{document}